# An LSB Data Hiding Technique Using Prime Numbers

Sandipan Dey [(1)], Ajith Abraham [(2)], Sugata Sanyal [(3)]

[1]Anshin Software Private Limited, Kolkata – 700091

[2]Centre for Quantifiable Quality of Service in Communication Systems
Norwegian University of Science and Technology, Norway

[3]School of Technology and Computer Science, Tata Institute of Fundamental Research, India
sandipan.dey@gmail.com, ajith.abraham@ieee.org, sanyal@tifr.res.in

## Abstract

*In this paper, a novel data hiding technique is proposed, as an improvement over the Fibonacci LSB data-hiding technique proposed by Battisti et al. [1]. First we mathematically model and generalize our approach. Then we propose our novel technique, based on decomposition of a number (pixel-value) in sum of prime numbers. The particular representation generates a different set of (virtual) bit-planes altogether, suitable for embedding purposes. They not only allow one to embed secret message in higher bit-planes but also do it without much distortion, with a much better stego-image quality, and in a reliable and secured manner, guaranteeing efficient retrieval of secret message. A comparative performance study between the classical Least Significant Bit (LSB) method, the Fibonacci LSB data-hiding technique and our proposed schemes has been done. Analysis indicates that image quality of the stego-image hidden by the technique using Fibonacci decomposition improves against that using simple LSB substitution method, while the same using the prime decomposition method improves drastically against that using Fibonacci decomposition technique. Experimental results show that, the stego-image is visually indistinguishable from the original cover-image.*

## 1. Introduction

Data hiding technique is a new kind of secret communication technology. While cryptography scrambles the message so that it can't be understood, steganography hides the data so that it can't be observed. In this paper, we discuss about a new decomposition method for classical LSB data-hiding technique, in order to make the technique more secure and hence less predictable. We generate a new set of (virtual) bit planes using our decomposition technique and embed data bit in these bit planes.

## 2. Fibonacci LSB Data Hiding Technique

The aim of this particular technique (proposed by Battisti et al) is to investigate decomposition into different bit-planes, based on Fibonacci–p-sequences,

$$F_p(0) = F_p(1) = 1$$

$$F_p(n) = F_p(n-1) + F_p(n-p-1), \forall n \geq 2, n \in \mathbb{N}$$

and embed a secret message-bit into a pixel if it passes the Zeckendorf condition, then during extraction, follow the reverse procedure.

## 3. A Generalized LSB Data Hiding and the Prime Decomposition Technique

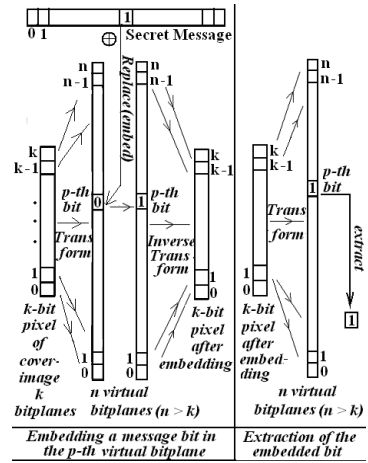

**Figure 1.** Generalized data-hiding technique

If we have k-bit cover image, only k bit-planes are available to embed secret data. Distortion increases exponentially with increasing bit-plane, it becomes impossible to embed data in higher bit-planes.

So, our primary target here is to increase the total number of available (and embeddable) bit planes without much distortion. To do this, we try to find a function $f$ that increases the number of bit-planes (for a k-bit image) from $k$ to $n, n \geq k$, by converting to some other binary number system with different weights, ensuring that number of bits taken to represent the same pixel is greater than that of classical binary (these extra bit-planes are referred to as virtual bit-planes), also ensuring less abrupt change in pixel value with increasing bit plane. It allows higher (virtual) bit planes to be used to embed data with much less distortion. Figures1 and 2 explain this concept.

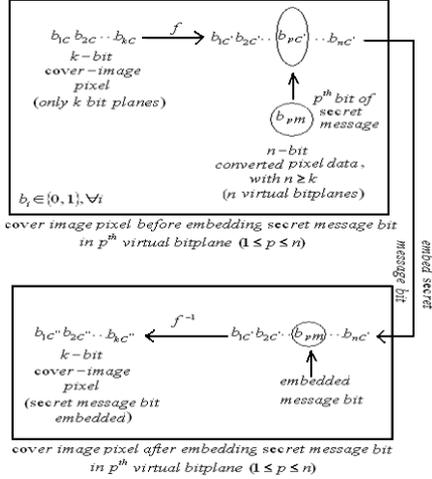

**Figure 2.** Illustration of embedding secret data-bit

### 3.1. The Number System
We define a number system by defining:
1. A constant, called base or radix '$r$' (digits of the number system $\in \{0,...,r-1\}$)
2. A function, called weight function $W(.)$, where $W(i)$ denotes weight corresponding to $i^{th}$ bit, $\forall i$

Hence, the pair $(r, W(.))$ defines a number system completely. A number having representation $d_{k-1}d_{k-2}..d_1d_0$ in number system $(r, W(.))$ will have value $D = \sum_{i=0}^{k-1} d_i.W(i), where, d_i \in \{0,1,..,k-1\}$ in decimal. Also, we may have more than one representation for the same number in our number system, we must be able to eliminate this redundancy and represent one number uniquely. We use the following strategy - from multiple representations of the same value, choose the one with lexicographical highest value, discard all others. For classical binary number system, we have,
$W(.) = 2^{(.)} \Rightarrow W : i \mapsto 2^i \Rightarrow W(i) = 2^i, \forall i \in Z^+ \cup \{0\}$,
corresponding to $i^{th}$ bit-plane ($LSB = 0^{th} bit$). A k-bit number (i.e. pixel-value) $p_k$ is represented as,
$p_k = \sum_{i=0}^{k-1} b_{iC}.2^i, where, b_{iC} \in \{0,1\}$. Now, $f$ converts this $p_k$ to some virtual pixel $p_n^/$ with n (virtual) bit-planes, $n \geq k$, to expand number of bit planes. To find such an $f$ is equivalent to finding a new weight function $W(.), i.e., W(i), \forall i \in \{0,..,n-1\}$, so that $W(i)$ denotes weight of $i^{th}$ virtual bit-plane in the new number system. $p_n^/ = \sum_{i=0}^{n-1} b_{iC}^/.W(i), b_{iC}^/ \in \{0,1\}$, our new decomposition, satisfying
$(p_k)_{(2,2^{(.)})} = (p_n^/)_{(2,w(.))}$

Also, $W(i)$ must have less abrupt changes with respect to increasing $i$ than that in case of $2^i$. Moreover, we must ensure that the function $f$ must be injective, i.e., invertible, otherwise we shall not be able to extract the embedded message precisely.

### 3.2. Number System Using Fibonacci p-Sequence Decomposition

The weight function proposed by Battisti et al. is $F_n, \forall n \in N$, i.e., $W(.) = Fib_p(.)$, number system to model virtual bit-planes is $(2, F_p(.))$. To ensure invertibility, instead of Zeckendorf's theorem, we prefer to use lexicographically higher property in case of Fibonacci as well, similar to what we shall use in case of our prime decomposition technique.

| N | Fib. Decomp | N | Fib. Decomp | N | Fib. Decomp | N | Fib. Decomp |
|---|---|---|---|---|---|---|---|
| 0 | 000000000000 | 32 | 000001010100 | 64 | 000100010001 | 96 | 001000001010 |
| 1 | 000000000001 | 33 | 000001010101 | 65 | 000100010010 | 97 | 001000010000 |
| 2 | 000000000010 | 34 | 000010000000 | 66 | 000100010100 | 98 | 001000010001 |
| 3 | 000000000100 | 35 | 000010000001 | 67 | 000100010101 | 99 | 001000010010 |
| 4 | 000000000101 | 36 | 000010000010 | 68 | 000100100000 | 100 | 001000010100 |
| 5 | 000000001001 | 37 | 000010000100 | 69 | 000100100001 | 101 | 001000010101 |
| 6 | 000000001001 | 38 | 000010000101 | 70 | 000100100010 | 102 | 001000100000 |
| 7 | 000000001010 | 39 | 000010001000 | 71 | 000100100100 | 103 | 001000100001 |
| 8 | 000000010000 | 40 | 000010001001 | 72 | 000100100101 | 104 | 001000100010 |
| 9 | 000000010001 | 41 | 000010001010 | 73 | 000100101000 | 105 | 001000100100 |
| 10 | 000000010010 | 42 | 000010010000 | 74 | 000100101001 | 106 | 001000100101 |
| 11 | 000000010100 | 43 | 000010010001 | 75 | 000100101010 | 107 | 001000101000 |
| 12 | 000000010101 | 44 | 000010010010 | 76 | 000101000000 | 108 | 001000101001 |
| 13 | 000000100000 | 45 | 000010010100 | 77 | 000101000001 | 109 | 001000101010 |
| 14 | 000000100001 | 46 | 000010010101 | 78 | 000101000010 | 110 | 001001000000 |
| 15 | 000000100010 | 47 | 000010101000 | 79 | 000101000100 | 111 | 001001000001 |
| 16 | 000000100100 | 48 | 000010101001 | 80 | 000101000101 | 112 | 001001000010 |
| 17 | 000000100101 | 49 | 000010100001 | 81 | 000101001000 | 113 | 001001000100 |
| 18 | 000000101000 | 50 | 000010100100 | 82 | 000101001001 | 114 | 001001000101 |
| 19 | 000000101001 | 51 | 000010100101 | 83 | 000101001010 | 115 | 001001001000 |
| 20 | 000000101010 | 52 | 000010101010 | 84 | 000101010000 | 116 | 001001001001 |
| 21 | 000001000000 | 53 | 000010101010 | 85 | 000101010001 | 117 | 001001001010 |
| 22 | 000001000001 | 54 | 000010101010 | 86 | 000101010010 | 118 | 001001010000 |
| 23 | 000001000010 | 55 | 000100000000 | 87 | 000101010100 | 119 | 001001010001 |
| 24 | 000001000100 | 56 | 000100000001 | 88 | 000101010101 | 120 | 001001010010 |
| 25 | 000001000101 | 57 | 000100000010 | 89 | 001000000000 | 121 | 001001010100 |
| 26 | 000001001000 | 58 | 000100000100 | 90 | 001000000001 | 122 | 001001010101 |
| 27 | 000001001001 | 59 | 000100000101 | 91 | 001000000010 | 123 | 001010000000 |
| 28 | 000001001010 | 60 | 000100001000 | 92 | 001000000100 | 124 | 001010000001 |
| 29 | 000001010000 | 61 | 000100001001 | 93 | 001000000101 | 125 | 001010000010 |
| 30 | 000001010001 | 62 | 000100001010 | 94 | 001000001000 | 126 | 001010000100 |
| 31 | 000001010010 | 63 | 000100010000 | 95 | 001000001001 | 127 | 001010000101 |

**Figure 3.** Fibonacci (1-sequence) decomposition for 8-bit image yielding 12 virtual bit-planes.

### 3.3. Proposed Prime Number Decomposition
We define a new number system, denoted as $(2, P(.))$, where the weight function $P(.)$ is defined as:
$P(0) = 1, P(i) = p_i, \forall i \in Z^+, p_i = i^{th} \Pr ime$,
$p_0 = 1, p_1 = 2, p_2 = 3, p_3 = 5,..$

Since the weight function here is composed of prime numbers, we name this number system as prime number system and the decomposition as prime decomposition. If any value has more than one representation in this number system, we always take the lexicographically highest of them, to assert invertible property. (e.g., the number 3 has 2 different representations in 3-bit prime number system, namely, 100 and 011, since we have,

$1.p_2 + 0.p_1 + 0.1 = 1.3 + 0.2 + 0.1 = 3$

$0.p_2 + 1.p_1 + 1.1 = 0.3 + 1.2 + 1.1 = 3$

But 100 is lexicographically (from left to right) higher than 011, we choose 100 to be valid representation for 3 in our prime number system and thus discard 011 as an invalid representation. $3 \equiv \max_{lexicogaphic}(100, 011) \equiv 100$.

Hence, the valid representations are:
$000 \leftrightarrow 0, 001 \leftrightarrow 1, 010 \leftrightarrow 2, 100 \leftrightarrow 3,$
$101 \leftrightarrow 4, 110 \leftrightarrow 5, 111 \leftrightarrow 6$

Now, we embed a secret data bit into a (virtual) bit-plane by simply replacing the corresponding bit by the data bit, only if we find that after embedding the resulting representation is a valid representation in our number system, otherwise we don't embed, just skip. This is only to guarantee the existence of the inverse function and correctness for extraction of our secret embedded message bit.

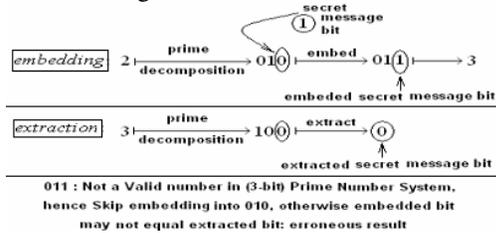

**Figure 4.** Error in not guaranteeing uniqueness
As evident from Figure-4, it's clear that one should embed secret data bit only to those pixels, where, after embedding, we get a valid representation in the number system.

### 3.4. Embedding Algorithm

First we find the set of all prime numbers that are required to decompose a pixel value in a k-bit cover-image, i.e., we need to find a number $n \in N$ such that all possible pixel values in the range $[0, 2^k - 1]$ can be represented using first n primes in our n-bit prime number system, so that we get 'n' virtual bit-planes after decomposition. Using Goldbach conjecture etc, that all pixel-values in the range $[0, \sum_{i=0}^{m-1} p_i]$ can be represented in our m-bit prime number system, so all we need to do is to find an 'n' such that $\sum_{i=0}^{n-1} p_i \geq 2^k - 1$, since the highest number that can be represented in n-bit prime number system is $\sum_{i=0}^{n-1} p_i$).

After finding the primes, we create a map of k-bit (classical binary decomposition) to n-bit numbers (prime decomposition), $n > k$, marking all the valid representations in our prime number system.

For an 8-bit image, part of pixel value vs. prime decomposition map is shown in Figure 5.

| N | Prime Decomp. | N | Prime Decomp. | N | Prime Decomp. |
|---|---|---|---|---|---|
| 0 | 000000000000000 | 32 | 000100000000001 | 64 | 100000100000010 |
| 1 | 000000000000001 | 33 | 000100000000010 | 65 | 100000100000100 |
| 2 | 000000000000010 | 34 | 000100000000100 | 66 | 100001000000000 |
| 3 | 000000000000100 | 35 | 000100000000101 | 67 | 100001000000001 |
| 4 | 000000000000101 | 36 | 000100000001000 | 68 | 100001000000010 |
| 5 | 000000000001000 | 37 | 001000000000000 | 69 | 100001000000100 |
| 6 | 000000000001001 | 38 | 001000000000001 | 70 | 100001000000101 |
| 7 | 000000000010000 | 39 | 001000000000010 | 71 | 100001000001000 |
| 8 | 000000000010001 | 40 | 001000000000100 | 72 | 100010000000000 |
| 9 | 000000000010010 | 41 | 010000000000000 | 73 | 100010000000001 |
| 10 | 000000000010101 | 42 | 010000000000001 | 74 | 100010000000100 |
| 11 | 000000000100000 | 43 | 100000000000000 | 75 | 100100000000001 |
| 12 | 000000000100001 | 44 | 100000000000001 | 76 | 100100000000010 |
| 13 | 000000001000000 | 45 | 100000000000010 | 77 | 100100000000100 |
| 14 | 000000001000001 | 46 | 100000000000100 | 78 | 100100000000101 |
| 15 | 000000001000010 | 47 | 100000000000101 | 79 | 100100000001000 |
| 16 | 000000001000100 | 48 | 100000000001000 | 80 | 101000000000000 |
| 17 | 000000010000000 | 49 | 100000000001001 | 81 | 101000000000001 |
| 18 | 000000010000001 | 50 | 100000000010000 | 82 | 101000000000010 |
| 19 | 000000100000000 | 51 | 100000000010001 | 83 | 101000000000100 |
| 20 | 000000100000001 | 52 | 100000000010010 | 84 | 110000000000000 |
| 21 | 000000100000010 | 53 | 100000000010100 | 85 | 110000000000001 |
| 22 | 000000100000100 | 54 | 100000000100000 | 86 | 110000000000010 |
| 23 | 000001000000000 | 55 | 100000000100001 | 87 | 110000000000100 |
| 24 | 000001000000001 | 56 | 100000001000000 | 88 | 110000000000101 |
| 25 | 000001000000010 | 57 | 100000001000001 | 89 | 110000000001000 |
| 26 | 000001000000100 | 58 | 100000001000010 | 90 | 110000000001001 |
| 27 | 000001000000101 | 59 | 100000001000100 | 91 | 110000000010000 |
| 28 | 000001000001000 | 60 | 100000010000000 | 92 | 110000000010001 |
| 29 | 000010000000000 | 61 | 100000010000001 | 93 | 110000000010010 |
| 30 | 000010000000001 | 62 | 100000100000000 | 94 | 110000000010100 |
| 31 | 000100000000000 | 63 | 100000100000001 | 95 | 110000000100000 |

**Figure-5.** Prime decomposition for 8-bit image yielding 15 virtual bit-planes

Next, for each pixel of cover image choose a (virtual) bit plane, say $p^{th}$ bit-plane $(p < n)$, embed secret data bit into that particular bit plane, by replacing the corresponding bit by the data bit, iff we find that after embedding the data bit, the resulting sequence is a valid representation in n-bit prime number system, i.e., exists in the map. After embedding the secret message bit, we convert the resultant sequence in prime number system back to its value (in classical binary number system) and get our stego-image.

The extraction algorithm is exactly the reverse. From stego-image, we convert each pixel with embedded data bit to its corresponding prime decomposition and from $p^{th}$ bit-plane extract secret message bit. Combine all bits to get the secret message.

## 3.5. Comparison Between Standard Binary, Fibonacci and Prime Decomposition

By Tchebychef theorem [5], we have,

$0.92 < \frac{\pi(x)\ln(x)}{x} < 1.105, \forall x \geq 2$,

where $\pi(x)$ = number of primes not exceeding x, which leads to the very famous Prime Number Theorem $\lim_{n \to \infty} \frac{\pi(n)}{(n/\ln(n))} = 1$. Now, from this, one can show that $\lim_{n \to \infty} \frac{p_n}{n \ln(n)} = 1$, if $p_n$ be the $n^{th}$ prime,

$\therefore p_n = \theta(n.\ln(n))$

### A Lower Bound for the Fibonacci Numbers

If $\alpha$ be a positive root of the quadratic equation $\alpha^2 - \alpha - 1 = 0$, i.e., $\alpha = \frac{1+\sqrt{5}}{2}$, it is easy to show (e.g., by mathematical induction) that,

$F(n) > \alpha^{n-1}, \forall n > 1, n \in N$. Since $\sqrt{5} \approx 2.236$, we get,

$F(n) > (1.618034)^{n-1}, \forall n > 1$

We can easily generalize the above definition of Fibonacci sequence into Fibonacci p-sequence,

$F_p(0) = F_p(1) = 1$

$F_p(n) = F_p(n-1) + F_p(n-p-1), \forall n \geq 2, n \in N$

For $p = 1$, we obtain Fibonacci 1-sequence, as defined above. Similarly, for other values of p, one can easily derive (by similar induction) some exponential lower-bounds, and it is quite obvious that the base of the exponential lower bound will decrease gradually with increasing p. e.g., for p = 2, if $\alpha$ be a positive root of the equation $\alpha^3 - \alpha^2 - 1 = 0$, solving (e.g., by Newton-Raphson) we get $\alpha = 1.465575$, and it's easy to show by induction that

$F_2(n) > (1.465575)^{n-1}, \forall n > 1$,

From above, we can generalize, for Fibonacci p-sequence, if $\alpha_p$ be a positive root of the equation $\alpha^{p+1} - \alpha^p - 1 = 0$, we have the inequality,

$F_p(n) > (\alpha_p)^{n-1}$,

$\alpha_p \in \Re^+, \alpha_1 = \frac{1+\sqrt{5}}{2} = 1.618034$,

$\alpha_2 = 1.465575, \alpha_3 = 1.380278, \alpha_4 = 1.324718$,

$\alpha_p > \alpha_{p+1}, \forall p \in Z^+$

The sequence $\alpha_p$ is decreasing in $p$.

## 3.6. Performance Measures

*Mean Squared Error and SNR:* We have the following test statistics for performance measures,

$MSE = \sum_{i=1}^{M} \sum_{j=1}^{N} (f_{ij} - g_{ij})^2 MN$

$PSNR = 10\log_{10}\left(\frac{L^2}{MSE}\right)$

where M and N are the number of rows and number of columns respectively of the cover image, $f_{ij}$ is the pixel value from the cover image, $g_{ij}$ is the pixel value from the stego-image, and L is the peak signal value of the cover image (for 8-bit images, L=255). Signal to noise ratio quantifies the imperceptibility, by regarding the message as the signal and the message as the noise. Here, we use a slightly different test-statistic, namely, Worst-case-Mean-Square-Error (WMSE) and the corresponding PSNR (per pixel) as our test-statistics. We define WMSE as follows:

If the secret data-bit is embedded in the $i^{th}$ bit plane of a pixel, the worst-case error-square-per-pixel

$= WSE = |W(i)(1-0)|^2 = W(i)^2$, the case when the corresponding bit in cover-image toggles in stego-image, after embedding the secret data-bit. (e.g., worst-case error-square-per-pixel for embedding in $i^{th}$ bit plane for a pixel in classical binary decomposition is $= (2^i)^2 = 4^i$. If the grayscale cover-image has size w x h, we define,

$WMSE = w \times h \times (W(i))^2 = w \times h \times WSE$. Here, we try to minimize this WMSE (hence WSE) and maximize the corresponding PSNR, where

$PSNR = 10\log_{10}\left(\frac{L^2}{WSE}\right)$

### 3.6.1. Proposed Prime Decomposition generates More (virtual) Bit-planes

Using Classical binary decomposition, for a k-bit cover image, we get only k bit-planes per pixel, where we can embed our secret data bit. Now, we have, $p_n = \theta(n.\ln(n))$ and

$\exists \alpha_p \in \Re^+ : F_p(n) > (\alpha_p)^{n-1}$

$n.\ln(n) = o(\alpha_p{}^n)$ directly implies $p_n = o(F_p(n))$

The maximum (highest) number that can be represented in n-bit number system using our prime decomposition is $\sum_{i=0}^{n-1} p_i$, and in case of $n$-bit number system using Fibonacci $p$-sequence decomposition is $\sum_{i=0}^{n-1} F_p(i)$. Now, it's easy to prove that $\exists n_0 \in N : \forall n \geq n_0$, we have,

$\sum_{i=0}^{n-1} F_p(i) > \sum_{i=0}^{n-1} p_i$. So, using same number of bits it is possible to represent more numbers in case of prime decomposition than in case of Fibonacci p-sequence decomposition, when number of bits is greater than some threshold. This in turn implies that number of (virtual) bit-planes generated in case of prime decomposition will be eventually (after some n) more than the corresponding number of (virtual) bit-planes generated by Fibonacci p-Sequence decomposition. Figure 6 illustrates this claim.

### 3.6.2. Prime Decomposition gives less distortion in higher bit-planes

Here we assume the secret message length (in bits) is same as image size, for evaluation of our test-statistics. For message with different length, the same can similarly be derived in a straight-forward manner. In case of our Prime Decomposition, WMSE for embedding secret message bit only in $l^{th}$ (virtual) bit-plane of each pixel (after expressing a pixel in our prime number system, using prime decomposition technique) $= p_l^2$, because change in $l^{th}$ bit plane of a pixel simply implies changing of the pixel value by at most $l^{th}$ prime number. From above, (treating image-size as constant) we conclude,

$\left(WMSE_{l^{th}bit-plane}\right)_{Prime-Decomposition} = w \times h \times p_l^2 = \theta(l^2 . \log^2(l))$.

whereas WMSE in case of classical (traditional) binary (LSB) data hiding technique is given by,

$\left(WMSE_{l^{th}bit-plane}\right)_{Binary-Decomposition} = \theta(4^l)$.

The above result implies that the distortion in case of prime decomposition is much less (polynomial) than for classical binary (exponential). Now, let's calculate the WMSE for the embedding technique using Fibonacci p-sequence decomposition. In this case, WMSE for embedding secret message bit only in $l^{th}$ (virtual) bit-plane of each pixel (expressing it using Fibonacci-1-sequence decomposition) $= (F_p(l))^2$, because change in $l^{th}$ bit plane of a pixel implies changing of pixel value by at most $l^{th}$ Fibonacci number. For $p = 1$,

$\left(WMSE_{l^{th}bit-plane}\right)_{Fibonacci-1-Sequence-Decomposition}$
$= w \times h \times (F(l))^2 = \theta((F(l))^2) > \theta((2.618)^l)$.

Similarly, for other values of p, one can easily derive (by induction) some exponential lower-bounds, which are definitely better than the exponential bound obtained in case of classical binary decomposition, but still they are exponential in nature, even if the base of the exponential lower bound will decrease gradually with increasing p. Generalizing, we get,

$\left(WMSE_{l^{th}bit-plane}\right)_{Fibonacci-p-Sequence-Decomposition} > \theta\left((\alpha_p^2)^l\right)$, T

$\alpha_p \in \Re^+, \alpha_1 = \frac{1+\sqrt{5}}{2}, \alpha_p^2 > \alpha_{p+1}^2, \forall p \in Z^+$

he sequence $\alpha_p^2$ is decreasing in $p$.

Obviously, the Fibonacci-p-sequence decomposition, despite being better than classical binary decomposition, is still exponential and causes much-more distortion in the higher bit-planes, than our prime decomposition, in which case WMSE is polynomial (and not exponential!) in nature.

The plot shown in Figure-6 proves our claim, it vindicates polynomial nature of the weight function in case of prime decomposition and exponential nature of classical binary and Fibonacci decomposition.

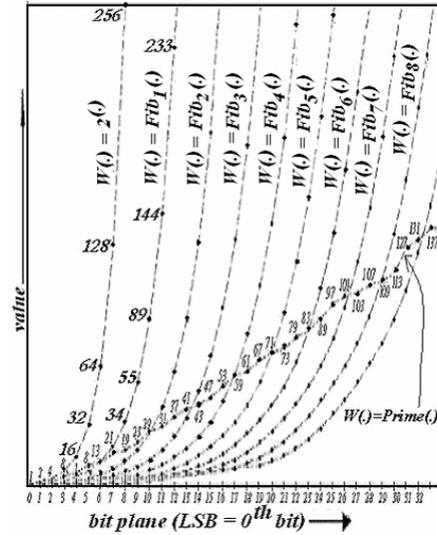

**Figure6.** Wt. functions for different decompositions

At a glance, the result of our test-statistics,
$\left(WMSE_{l^{th}bit-plane}\right)_{Classical-Binary-Decomposition} = \theta(4^l)$,
$\left(WMSE_{l^{th}bit-plane}\right)_{Prime-Decomposition} = \theta(l^2 . \log^2(l))$,
$\left(WMSE_{l^{th}bit-plane}\right)_{Fibonacci-p-Decomposition} = \theta((c_p)^l)$,
$c_p \in \Re^+, 2.618 > c_p > c_{p+1}, \forall p \in Z^+, with$
$\left(WMSE_{l^{th}bit-plane}\right)_{Fibonacci-1-Decomposition} = \theta((2.618)^l)$.

$$(PSNR_{worst})_{Classical-Binary-Decomposition} = 10.\log_{10}\left(\frac{(2^k-1)^2}{(2^l)^2}\right)$$

$$(PSNR_{worst})_{Prime-Decomposition} = 10.\log_{10}\left(\frac{(2^k-1)^2}{(c.l^2.\log^2(l))^2}\right), c \in \Re^+$$

$$(PSNR_{worst})_{Fibonacci-p-Decomposition} = 10.\log_{10}\left(\frac{(2^k-1)^2}{(c_p)^l}\right),$$

$$\alpha_p \in \Re^+, \alpha_1 = 2.618, \alpha_p > \beta_{p+1}, \forall p \in Z^+, with$$

$$(PSNR_{worst})_{Fibonacci-1-Decomposition} = 10.\log_{10}\left(\frac{(2^k-1)^2}{(2.618)^l}\right)$$

## 4. Experiment Results

We have, as input, an 8-bit gray-level cover image of Lena. Secret message length = cover image size, (message string 'sandipan' repeated multiple times to fill the cover image size). The secret message bits are embedded in chosen bit-plane '$p$'. The test message is hidden into the chosen bit-plane using the classical binary (LSB) technique, Fibonacci (1-sequence) decomposition and Prime decomposition technique separately and compared.

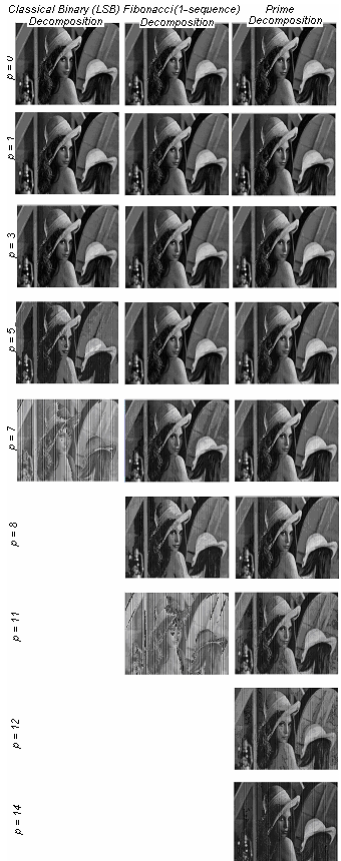

**Figure 7**. Results of embedding data in different bit-planes using different data-hiding techniques

Figure 7 illustrates that we get 8, 12 and 15 (virtual) bit-planes using classical LSB, Fibonacci and Prime decomposition data-hiding technique respectively (highest 15 virtual bit-planes for Prime). Data-hiding technique using the prime decomposition has a better performance than that of Fibonacci decomposition, the later being more efficient than classical binary decomposition, when judged in terms of embedding secret data bit into higher bit-planes causing least distortion, thereby least chance of being detected. To embed in more than one virtual bit-plane, one may use variable depth embedding [2].

## 5. Conclusions

This paper presented very simple method of data hiding technique using prime numbers. It is shown (both theoretically and experimentally) that the data-hiding technique using prime decomposition outperforms the famous LSB data hiding technique using classical binary decomposition, and that using Fibonacci p-sequence decomposition. We have experimented using the famous Lena image, but since our theoretical derivation illustrates that the test-statistic value (WMSE, PSNR) is independent of the probability mass function of the gray levels of the input image, the (worst-case) results will be similar if we use any gray-level image as input, instead of the Lena image.